# Bio-inspired Compact, High-resolution Snapshot Hyperspectral Imaging System with 3D Printed Glass Lightguide Array

ZHIHAN HONG,[1, †] YUANYUAN SUN,[1, †] PIAORAN YE,[1, †] DOUGLAS A. LOY,[2,3] AND RONGGUANG LIANG[1,*]

[†] These authors contributed equally to this work
[1] Wyant College of Optical Sciences, The University of Arizona, 1630 E University Blvd, Tucson, AZ 85721, USA
[2] Department of Chemistry & Biochemistry, The University of Arizona, 1306 E. University Blvd, Tucson, AZ 85721-0041, USA
[3] Department of Materials Science & Engineering, The University of Arizona, 1235 E. James E. Rogers Way, Tucson, AZ 85721-0012, USA
*rliang@optics.arizona.edu

## Abstract

To address the major challenges to obtain high spatial resolution in snapshot hyperspectral imaging, 3D printed glass lightguide array has been developed to sample the intermediate image in high spatial resolution and redistribute the pixels in the output end to achieve high spectral resolution. Curved 3D printed lightguide array can significantly simplify the snapshot hyperspectral imaging system, achieve better imaging performance, and reduce the system complexity and cost. We have developed two-photon polymerization process to print glass lightguide array, and demonstrated the system performance with biological samples. This new snapshot technology will catalyze new hyperspectral imaging system development and open doors for new applications from UV to IR.

## Introduction

Snapshot hyperspectral imaging is an emerging imaging modality to obtain spectral information of the dynamic target in real time. Several snapshot hyperspectral imaging technologies have been developed for various applications ranging from astronomy to agriculture and from surveillance to medicine.[1-4] These technologies include integral field imaging, image mapping, multi-aperture filtering, spectrally resolving detector array, coded aperture, computed tomography imaging, tunable echelle imager, and Fourier transform spectral imaging.[1] Compared to other methods, integral field imaging method has a number of key advantages, such as easy to implement with regular camera, less computation, and low cost.

The key in integral field imaging technology is how to sample the intermediate field. Figure 1(a) is the schematic configuration of a typical snapshot hyperspectral imaging system. The imaging optics images the object to the input end of the field sampler which redistributes the image points to different locations in the output end. The collimation optics then collimates the light from the re-sampled image points. After passing through the dispersion element, the dispersed collimated light is focused onto the digital sensor. The hyperspectral images can then be reconstructed from the dispersed image. Pinhole array, faceted mirror, fiber bundle, and lenslet array have been used to sample the field, each of them has its unique properties and limitations.[5-10] Pinhole array method is simple and low cost to implement, but light efficiency and spatial resolution are two major limitations, preventing its adoption in practical applications. Lenslet array approach partially addresses the light efficiency issue, but the spatial resolution is still the bottleneck. Faceted mirror method is relatively difficult to implement as it is difficult to fabricate and the edges of the faceted mirrors can cause some artifacts. Fiber bundle approach is very straightforward, with one end of the fiber array arranged as compact as possible to sample the intermediate image and the other end of the fiber array specially arranged so that the spectral information can be distinguished in the detection sensor with a dispersion element, such as prism or grating. The resolution is determined by the pitch of the fiber bundle in the input end. Due to the cladding layer, the spatial resolution is still sacrificed and the light efficiency is not high either.

Three-dimensional (3D) printing is an emerging fabrication method for precision optics, it is attractive due to its flexibility in building complex shapes through an *additive* process.[11-22] Most of the research in printing optics to date has focused on organic polymer or resin-based systems using stereolithography (STL),[13] direct ink writing (DIW),[14] projection microstereolithography (PμSL),[15,16] and two-photon polymerization (TPP) processes.[19-22] Optical element printed with organic polymer or resin through UV curing process has a number of limitations in hardness, transparency in UV and NIR, thermal resistance, chemical resistance and tenability. Another issue is that the printed part becomes yellowish gradually, the transmission in short wavelength is further degraded. Compared to optics made from optical polymers and optical silicones, inorganic glass optics is preferred for many applications because of its excellent optical, chemical, and thermal properties. For the applications in UV and IR, glass optics is the only solution.

We have developed a compact, high-resolution snapshot hyperspectral imaging (HSHI) system with 3D printed glass lightguide array as shown in Figure 1(a). The imaging optics images the object onto the intermediate image plane, and the input end of the 3D printed lightguide array samples the intermediate image in high resolution since each lightguide can be printed as small as the pixel size of a typical digital sensor. The output end of the lightguide array is specially arranged so that the light spectrum from each lightguide, after passing through the collimation optics, dispersion element, and focusing optics, can be disentangled in the sensor with minimum crosstalk. In addition to the high spatial resolution, the proposed HSHI system can be very compact since the input and output surfaces of the lightguide array can be printed in freeform shape, including concave surfaces. With concave surfaces, the optical systems can be simplified dramatically as there are no needs to correct field curvature, which is one of the driving factors for complex optical system.[23-26]

In this paper, we will first introduce the concept of compact, high-resolution snapshot hyperspectral imaging with 3D printed glass lightguide array, and then demonstrate the system performance with mask pattern and biological slides. We will also discuss the future research and potential applications.

**Results**

**Concept of lightguide array for HSHI system**

The goal of snapshot hyperspectral imaging is to obtain spectral information of an object with a single capture. To obtain high spatial and spectral resolution, due to the two-dimensional (2D) structure of the digital sensor it is necessary to re-arrange the images of object points in the image plane so that their positions are sparsely located in the digital sensor. The pixels between adjacent lightguides are reserved for capturing spectral information.

Inspired by the unique features of vertebrate and arthropod eyes, we have developed lightguide array to address the major hurdle in snapshot hyperspectral imaging. Figure 1(b) is the schematic longitudinal section of an anterior median eyes of the funnel-web spider Agelena labyrinthica.[27] Below the cuticular lens lie the vitreous cells; they are enclosed by a cup of pigment cells. Similar to the human eye, the retina is curved to cover a larger field of view (FOV) and obtain better image quality. The retina consists of slim photoreceptor cells under a thin layer of pigment-free pigment cell processors. The light absorbing rhabdomeres extend on the distal sides of each visual cell. The axons of adjacent sensory cells meet near the middle of the eye cup. Figure 1(c) is the schematic structure of an apposition eye.[28] In most apposition eyes the eight receptor cells in each ommatidium contribute to a single radial structure, known as a rhabdom. Having a slightly higher refractive index than its surroundings, the rhabdom behaves as a lightguide, so that the light entering its distal tip travels down the structure, trapped by total internal reflection (TIR). Inspired by the unique features of the eyes in Figures 1(b) and 1(c), we have developed various lightguide arrays for hyperspectral imaging. Figure 1(d) is the schematic configuration of a lightguide array with lenses and dispersion prism in the output end. An imaging optical system is needed to image the object to the flat input surface of this lightguide array. The dispersed light from the output end will be detected by the sensor. Figure 1(e) is the schematic configuration of another lightguide array with imaging lens in each lightguide in the input end, lenses and dispersion prism in the output end to obtain the spectral information. This apposition eye configuration will enable the lightguide array to capture information from large FOV and obtain hyperspectral information from the flat sensor in the output end.

Figures 1(f)-1(h) are the schematic structures of the lightguide arrays with a small number of lightguides to transform the spatial distribution of the intermediate image formed by the optical imaging system to the optimal distribution for obtaining spectral information with minimum crosstalk. The lightguide array in Figure 1(f) has a concave input surface and convex output surface, mimicking the eye structure in Figure 1(b). The concave input surface can simplify the imaging optics as there is no need to correct field curvature. The convex output surface can display the information at large angles. The lightguide array in Figure 1(g) has concave surfaces on both ends, significantly simplifying both the imaging optics in front of the lightguide array and collimation optics after the lightguide array. The lightguide array in Figure 1(h) has flat surfaces on both ends, it is much simpler, suitable for the imaging optics and collimation optics designed for the flat field.

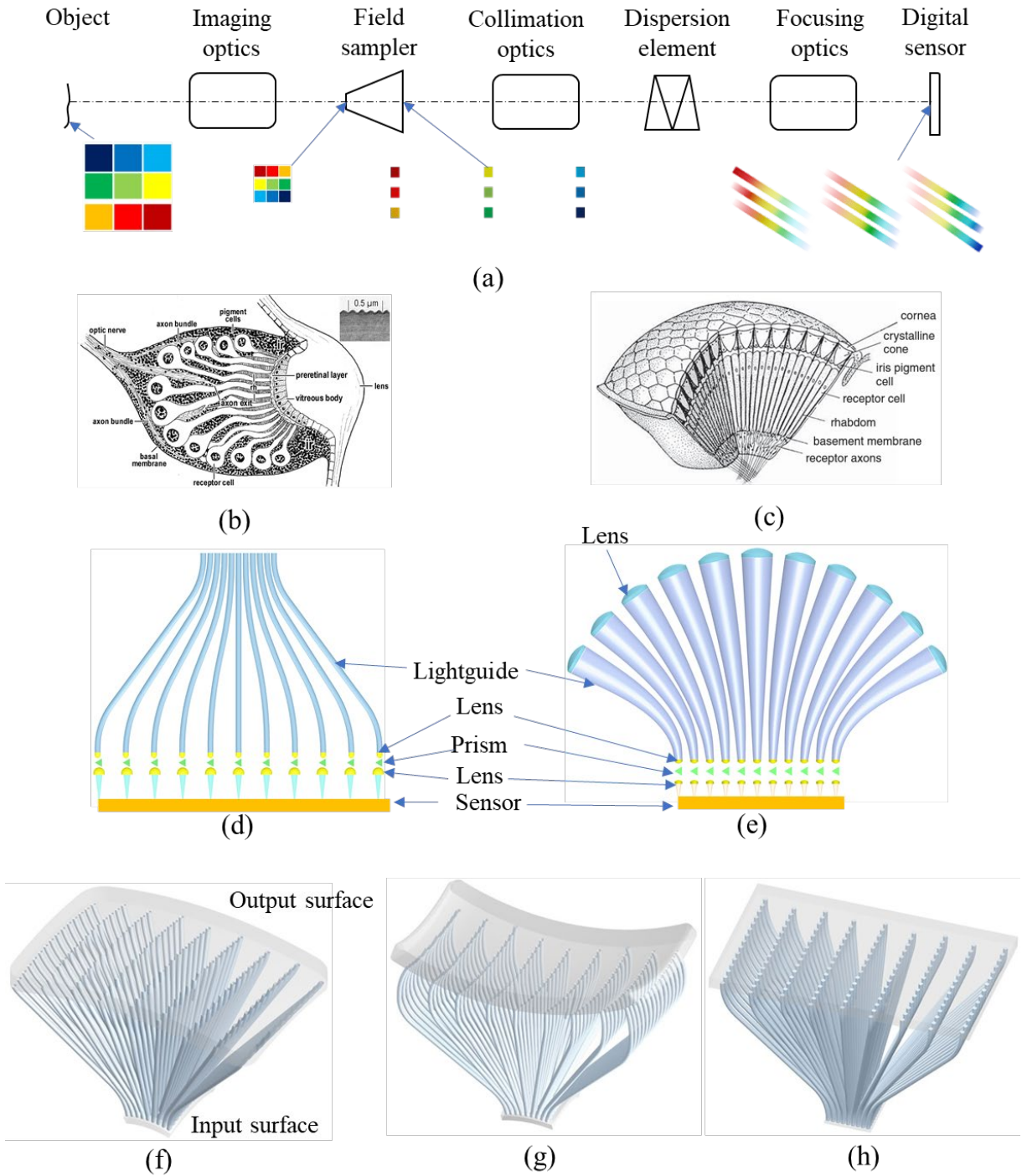

**Figure 1**. Bio-inspired compact, high-resolution snapshot hyperspectral imaging (HSHI) system. (a) Schematic configuration of HSHI with lightguide array, (b) eye structure of funnel-web spider Agelena labyrinthica,[26] (c) schematic diagram of an apposition eye,[27] (d) lightguide array with lenses and dispersion prism in the output end for hyperspectral imaging, (e) lightguide array with lens in the input end and lenses and prism in the output end for hyperspectral imaging, (f) lightguide array with concave input surface and convex output surface, (g) lightguide array with concave input and output surfaces, and (h) lightguide array with flat input and output surfaces.

The function of the lightguide array is to transform the spatial distribution of pixels from continuous mode in the intermediate image plane to sparse mode in the output end as shown in Figure 1(a). The lightguides in the input end are arranged in grid format, like the pixel layout in the monochromatic digital sensor. This allows the intermediate image to be sampled with high spatial resolution, and the hyperspectral image can be simply reconstructed in grid format from the dispersed image. Figures 2(a) and 2(b) are two layouts of the input ends of the lightguide arrays with square and circular lightguides. Square-shape lightguide has a better light efficiency and less light leakage. In contrast, the gaps between circular lightguides will leak light, reducing image contrast and introducing errors in reconstructing hyperspectral image. The lightguide shape can be maintained from the input end to the output end, the lightguide arrays in Figures 1(g)-1(h) and 2(c) are some examples. The shape can also change gradually from one shape to another, for example the lightguide in Figure 2(d) transforms the shape from square in the input end to circular in the output end. The arrangement of the lightguide array in the output surface is optimized to efficiently utilize the pixels in the detector based on the required spectral resolution.

Since the output surface of the lightguide is typically much larger than the input surface, all but the central lightguide are curved to some degrees, depending on the locations in the input and output ends as shown in Figure 2(e). One design guideline is to maintain the TIR condition along the entire lightguide. For the lightguide arrays with a large number of lightguides, the height $H$ in Figure 2(e) should be large enough to ensure that the local radius $R$ of the curve is large enough to avoid light leakage for the lightguide with large $W$.[29] Another design consideration is that the lightguide in both input and output ends should be normal to the input and output surfaces, regardless of being flat or curve, as shown in Figure 2(e). This ensures high light efficiency in coupling light into the lightguide array from the imaging system and coupling light out of the lightguide array for the collimation optics. The diameter of the individual lightguide should be small enough, and the fill-factor in the input end of the lightguide array should be high enough to achieve high spatial resolution imaging. Printed lightguide will be an ideal solution as there is no need to have the cladding layer.

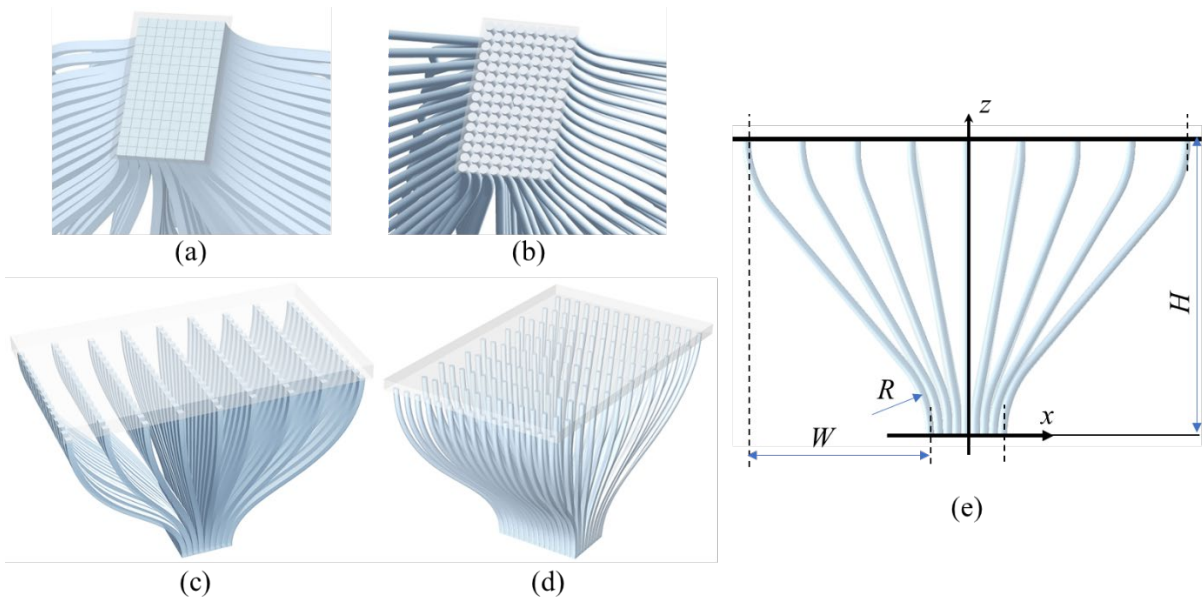

**Figure 2.** Design consideration of lightguide array for HSHI. (a) Layout of the square lightguide array in the input surface, (b) layout of the circular lightguide array in the input surface, (c) lightguide array with square lightguides, (d) lightguide array with lightguide gradually changing the shape from square in the input end to circular in the output end, and (e) design considerations of lightguide array.

**Printed lightguide array**

3D printing is attractive due to its flexibility in building complex shapes through an additive process.[30] Most reported studies use polymeric materials to print organic optics with decent performance. However, organic optics printed by polymer-based components are limited in practical applications due to their poor thermal stability, low transmission in short wavelengths, and low refractive indices. Glass optics is preferred as it has better optical, chemical, thermal, and mechanical properties. In the past few years, we have developed an TPP printing process with a solvent-free, pre-condensed liquid silica resin (LSR) for fabricating micro glass optics with relatively complex structures.[19,20] Transparent glass optics can be obtained after thermal treatment at 600 °C in the air with linear shrinkage as low as 17%. The peak-to-valley deviation of the printed surface is better than 100 nm and the surface roughness is better than 6 nm.[19] The solvent-free LSR is synthesized based on acid-catalyzed polymerization of tetramethoxysilane (TMOS) together with a sub-stoichiometric amount of water (water solution) and 6.5 mol% of methacryloxymethyltrimethoxysilane (MMTS) as photocurable moiety. Figure 3 shows the TPP printing process with LSR, Figure 3(a) is the schematic diagram of focusing the laser light to a small spot to cure LRS locally in the focal point and Figure 3(b) illustrates the pyrolysis process of the cured structure. Figure 3(c) is the SEM image of the printed micro-spectrometer consisting of a prism and a collimation lens with grating on the flat surface, and Figure 3(d) is the SEM image of a printed micro-objective.[20]

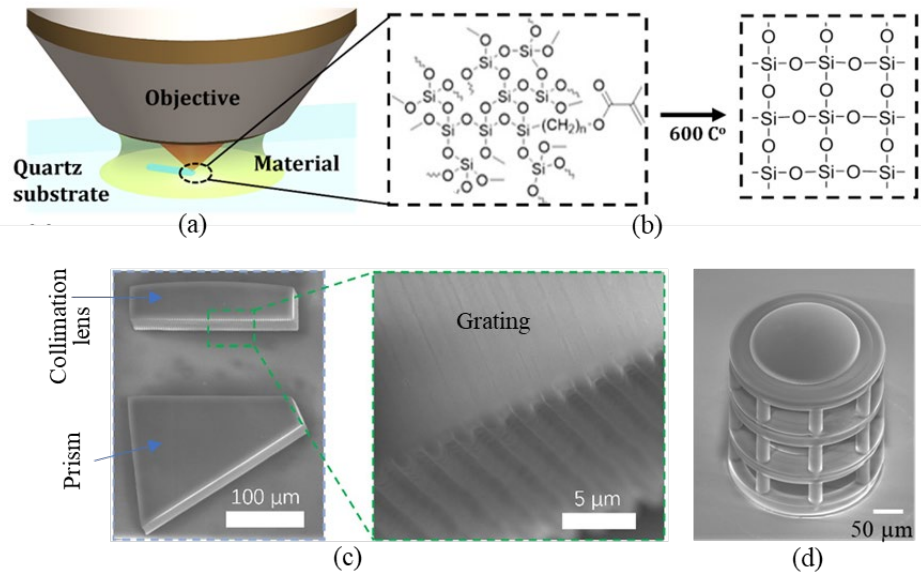

**Figure 3**. 3D printing process of glass optics. (a) Schematic diagram of the curing process, (b) the pyrolysis process of the cured structure, (c) printed micro-spectrometer consisting of a prism and a collimation lens with grating printed on the flat surface, and (d) printed micro-objective.

To demonstrate the capability of the printing system and materials, we have printed lightguide arrays with circular and square lightguides, different input and output surface shapes. Figure 4(a) is the SEM image of the printed lightguide array with convex input and output surfaces, Figure 4(b) is the SEM image of the 1/4 cross-section of the printed lightguide array with concave input and output surfaces, and Figure 4(c) is the SEM image of the lightguide array with flat input and output surfaces. Figures 4(d)-4(f) show the details of lightguide arrays around the input end, output end, and middle region, demonstrating smoothness of the lightguide TIR surface. All lightguide arrays in Figure 4 have the supporting structures to support the input and output surfaces.

Due to the limitations of the home-built 3D TPP printer without high precision motion stages, it is challenging to print the lightguide array with large height $H$, limiting its capability in printing square shape lightguides without breaking TIR condition because it is necessary to bend in two-dimension for the lightguides not in *x-z* and *y-z* planes. In this paper, the lightguide arrays with circular lightguides were printed to demonstrate the HSHI concept and evaluate system performance. With long-range 3-axis high precision motion stages, it is feasible to print the proposed lightguide array having more than 640x480 lightguides in either square or circular shape.

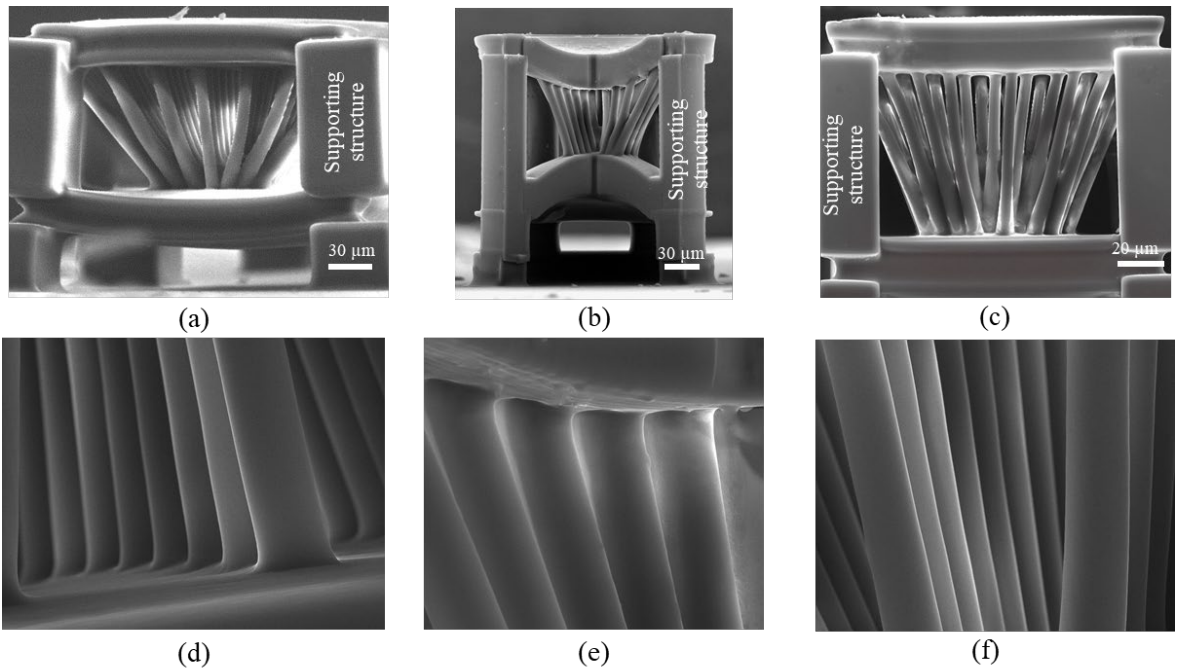

**Figure 4**. Printed lightguide arrays. SEM images of the lightguide array with (a) convex input and output surfaces, (b) concave input and output surface, and (c) flat input and output surface. SEM images of the individual lightguide around (d) the input surface, (e) output surfaces, and (f) middle region.

**Characterization of the lightguide array**

To demonstrate the concept with commercially available microscope objective and other imaging optics designed for flat object and image planes, the lightguide array with circular lightguides in Figure 1(h) was chosen for evaluation. Due to the limited printing volume of the home-built 3D TPP printer, a lightguide array with 15x9 lightguides was printed. The diameter of each lightguide was 5 µm, and the distance between the input and output surfaces was 100 µm. To measure the transmission of the lightguide, the light was first focused to a small spot and then coupled into the lightguide as shown in Figure 5(a). The ratio of the light received by the sensor with and without the lightguide is the transmission of the light guide. The light loss includes the light loss from Fresnel reflection on both surfaces, light leakage due to printing defects which break TIR condition, and material absorption. By changing the light spectrum, we measured the transmission of the lightguide at different wavelengths. Figure 5(b) plots the averaged transmission of the 9 lightguides.

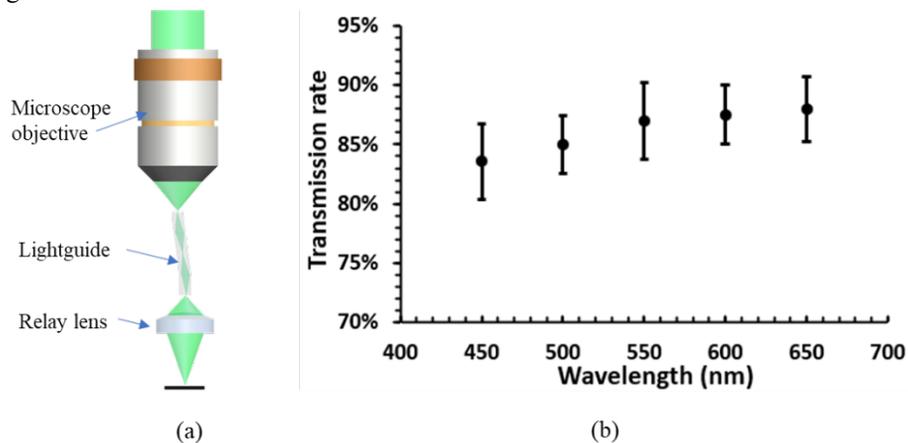

**Figure 5**. Transmission of the printed lightguide array. (a) Schematic diagram for measuring lightguide transmission, and (b) the averaged transmission of 9 lightguides.

To evaluate the performance of the 3D printed glass lightguide array in snapshot hyperspectral imaging, we built the system in Figure 6. A broadband light source (NKT Photonics FIU-15) provided illumination on the object with the help of the illumination optics (L1 and L2). The object, a Bayer filter segment, was relayed onto the input surface of the lightguide array using two microscope objectives (OBJ1: OLYMPUS UPlanFLN NA0.13 and OBJ2: NeoDPlan NA0.4). A third microscope objective (OBJ3: OLYMPUS LUCPlanFLN NA0.60) imaged the intermediate image from the output surface of the lightguide array to infinity. The collimated light was dispersed by the custom dispersion prism and then focused onto the digital sensor (IDSUI-3080SE-C) with the $f$=200 nm imaging lens (L3). The hyperspectral images were decoded from the dispersed image directly, and color image was reconstructed as superposition of the hyperspectral images according to the quantum efficiency of the color CMOS sensor.

The HSHI system integrates the spectral dimension with the spatial coordinates according to the lightguide structure. It encodes a 3D hypercube $HS(x, y, \lambda) \in \mathbb{R}^3$ into a 2D dispersed image $I^{disp}(x', y') \in \mathbb{R}^2$ as

$$f: (x, y, \lambda) \to (x', y') \quad (1)$$

where $f$ denotes the encoding mapping from the spatial-spectral coordinates $(x, y, \lambda)$ to the pixel coordinates $(x', y')$ on the sensor. To retrieve the hypercube, a decoding operation is necessary. Typically, the decoding operation consists of spatial and spectral calibrations. Spatial calibration decides the mapping between lightguide index $(x, y)$ and pixel coordinates on the dispersed image $(x', y')$.

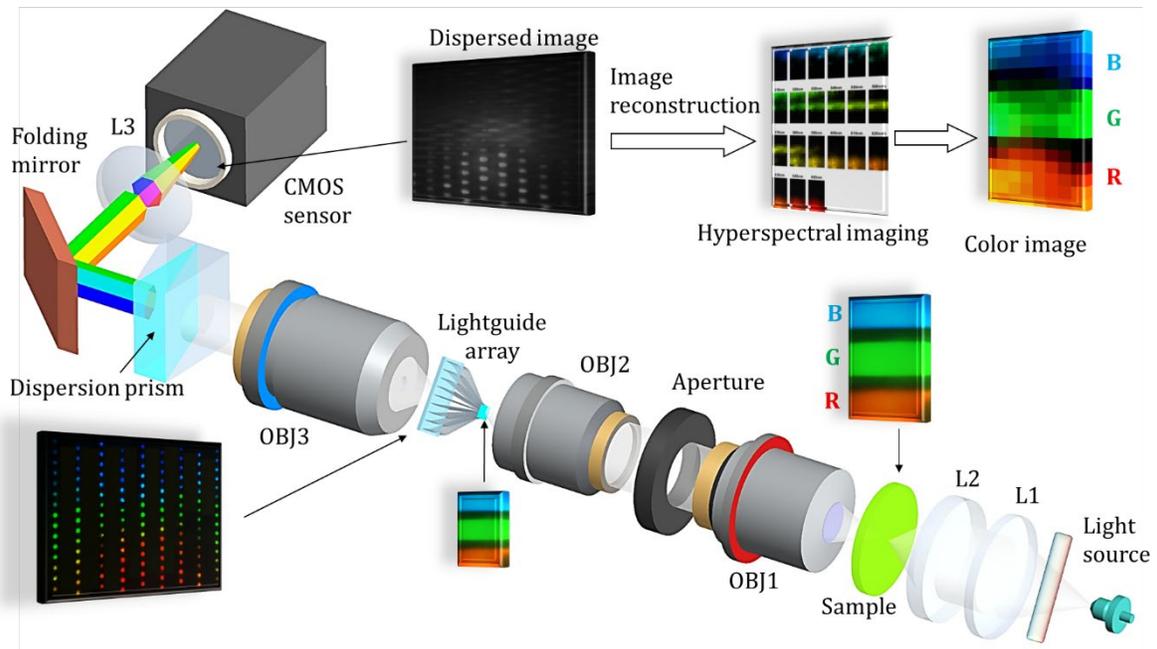

**Figure 6**. Experimental setup for evaluating the performance of the proposed HSHI system.

**Spatial calibration**

Spatial calibration intends to determine the pixel coordinates on the dispersed image for every lightguide in the input end. Since the relative position of the lightguide array in the input and output ends was known as shown in Figures 7(a) and 7(b), the even columns in the output end were shifted half period to keep interleaved with the odd columns. This interleaved pattern allowed sufficient pixels to cover the dispersed light from each lightguide and made full use of the pixels on the sensor without spectral overlap as shown in Figure 7(c) captured under visible illumination without the sample in the object plane (this image was captured with color CMOS sensor for illustration only). The circles in Figures 7(a) and 7(b) indicated the input end and output end of the same lightguide, the rectangular box in

Figure 7(c) was the dispersed spectrum of the light from the corresponding lightguide in Figure 7(b). The pixel coordinate for each lightguide was recorded for spatial calibration.

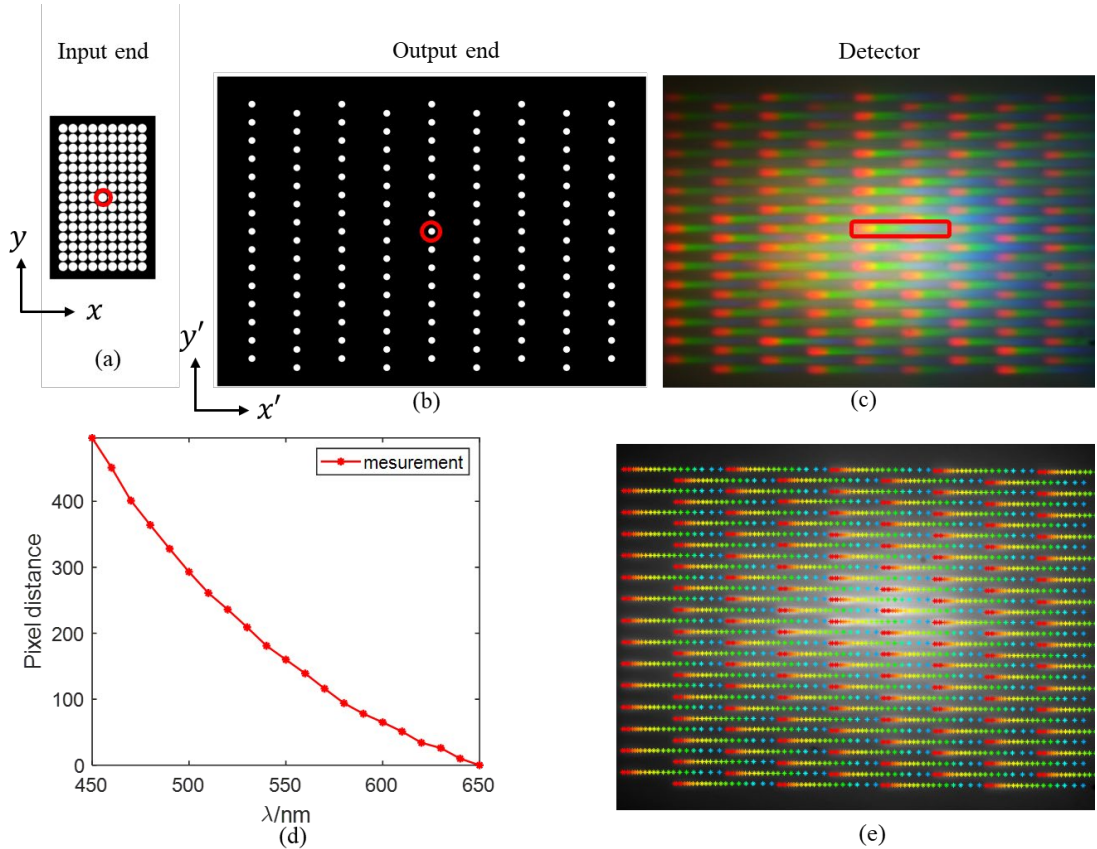

**Figure 7**. Spatial and spectral calibration. (a) Layout of lightguides in the input end, (b) layout of the lightguides in the output end, (c) dispersed image captured with color CMOS sensor, (d) the pixel distance from reference band (640-650nm) to other wave bands from 450nm to 650nm, and (e) the pixel coordinates for 21 spectral bands for 15 × 9 lightguides, the color of marker varies according to the corresponding wavelength.

**Spectral calibration**

The purpose of spectral calibration is to locate the spectral position on the dispersed image for each lightguide. To obtain accurate spectral location, an NKT SuperK EXTREME supercontinuum white light laser was used to provide illumination step-by-step from 450nm to 650nm with a 10nm increment. The corresponding spectral image was captured to pin the central location for each wavelength band. The pixel distances from the central pixel of 640-650nm waveband to the rest wavebands were calculated and plotted in Figure 7(d). The total pixel distance of the whole wavelength range was approximately 496 pixels, leaving a gap of around 42 pixels between the head and tail of the dispersed spectra adjacent lightguides. It made efficient use of pixels between neighboring light guides and effectively prevented crosstalk. After spatial and spectral calibration, the central pixel coordinates for 15 × 9 light guides at 21 wavebands were obtained and plotted in Figure 7(e) with the marker's color matching the wavelength.

**Hyperspectral reconstruction**

To retrieve the transmittance/reflectance signature of the object, flat-field correction is required to exclude the effect of illumination non-uniformity and the defects in the printed lightguides. In practice, an unload image $I^X$ under the same illumination without the sample was captured pair-wisely with the dispersed image $I^O$ with the sample. Two corresponding hypercubes $HS^X$ and $HS^O$ were then obtained after applying the spatial and spectral calibration on the two dispersed images. Their relationship can be described as Eqs. (2) and (3).

$$HS^O(x, y, \lambda) = i^{WL}(x, y, \lambda) \cdot T^O(x, y, \lambda) \cdot T^{LG}(x, y, \lambda) \qquad (2)$$

$$HS^X(x, y, \lambda) = i^{WL}(x, y, \lambda) \cdot T^{LG}(x, y, \lambda) \qquad (3)$$

$T^O$ and $T^{LG}$ represent the transmittance of the sample and light guide separately, and $i^{WL}$ denotes the white light illumination distribution. In experiments, both illumination and lightguide experienced certain inhomogeneities and it was difficult to isolate them from each other. Flat-field correction compensated for such illumination and system non-uniformity by dividing $HS^O$ by $HS^X$ as described in Eq. (4).

$$T^O(x, y, \lambda) = \frac{HS^O(x, y, \lambda)}{HS^X(x, y, \lambda)} \qquad (4)$$

Besides non-uniformity, another factor which degraded the reconstruction quality was noise. The noise mainly came from the light leakage from the gaps between the circular lightguide in the input end and the leakage during propagating through the lightguide due to the local surface defects. In addition, flat-field correction might boost the noise especially in areas with low intensity. A background noise term was removed for each lightguide spectral profile to reduce the impact of light leakage. The background noise was determined experimentally based on the noise in the surrounding black area. A thresholding operation was then applied to reduce the noise caused by flat-field correction, which appeared as large singular values. This strategy thresholded these singular values according to the spectral information of the neighboring lightguides.

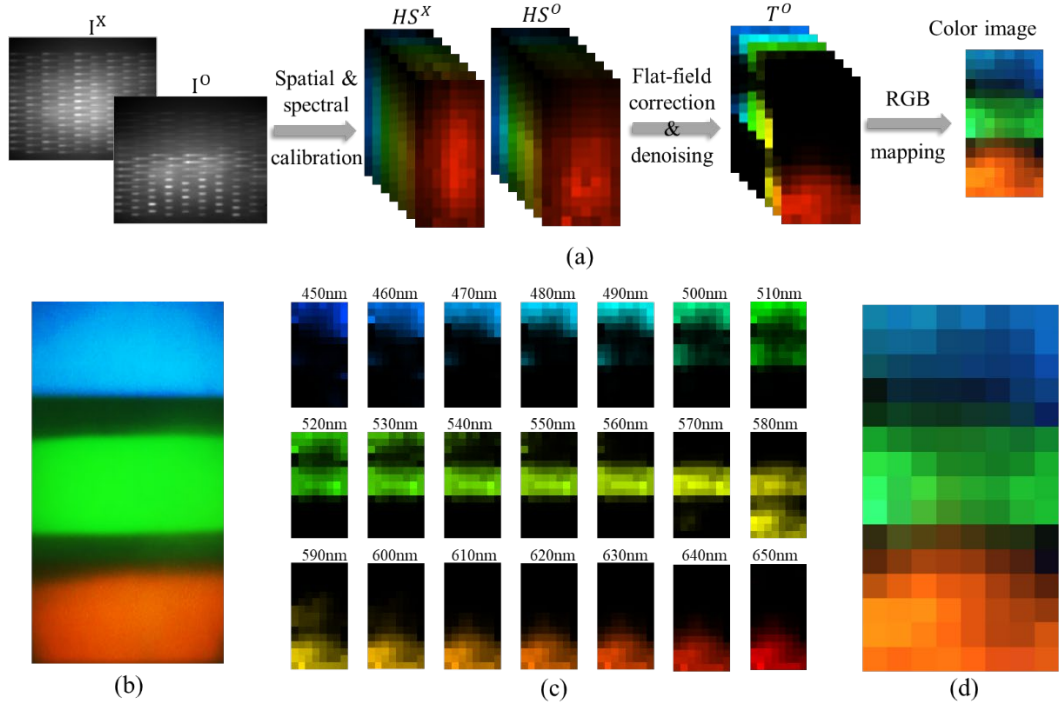

**Figure 8**. Image reconstruction pipeline. (a) Image reconstruction pipeline of proposed HSHI system with a Bayer filter segment, (b) the color image of a Bayer filter segment captured from a color sensor by removing the lightguide array and dispersion prism, (c) hyperspectral images from 450 nm – 650nm with 10 nm bandwidth, and (d) the reconstructed color image from the hyperspectral images in (c).

Figure 8(a) illustrates the image reconstruction pipeline with a Bayer filter sample in Figure 8(b). After spatial and spectral calibration, two 2D dispersive images $I^X$ and $I^O$ were decoded into two 3D hyperspectral cubes $HS^X$ and $HS^O$. With flat-field correction and denoising, $T^O$ exhibited a clear partitioning of the three filter bars, as demonstrated by the hyperspectral images with 10nm bandwidth in Figure 8(c). From the hyperspectral images, an RGB color image was then reconstructed as shown in Figure 8(d) with the quantum efficiency of the CMOS camera used to capture Figure 8(b). Compared to the color image captured without the lightguide array and the dispersion prism, the HSHI outcome exhibited a general resemblance with few differences due to the imperfections of the 3D printed lightguide array. The first issue was that the number of lightguides was insufficient, leading to the pixelated structure. The second issue was the light leakage in the central portion of the lightguide array, which could be clearly observed from $I^X$ where the central region was brighter, and the contrast was lower. The light leakage was due to the fact that the input end of the lightguide was circular, not square, leaving gaps between lightguides. This issue could be solved by printing the lightguide with square input end so that the fill factor is close to 100%, regardless of square or round output end. The length of the lightguide needs to be long enough so that the radius of curved region of the lightguide is large enough to avoid the light leakage during light propagating inside the lightguide.

**Experimental results**

To demonstrate the concept and validate imaging performance, a manmade flower with a red flower petal and a green stem (Figure 9(a)) was imaged first. The red petal and green stem were made from the transparent color film lighting gel filters. Figure 9(b) is the dispersed image captured by the camera, the extracted hyperspectral images as well as the reconstructed color image are illustrated in Figure 9(c) and 9(d) respectively. The hyperspectral images at each wavelength band demonstrated the variation of the spectral response at different locations. The reconstructed color image displayed a good consistency compared with the ground truth image captured by a color CMOS camera, considering under-sampling issue due to the limited number of lightguides. To validate the imaging performance, the measured spectral profiles of the light from two lightguides, which are two pixels in Figure 9(d), are plotted and compared with the transmitted spectral profiles of the color film filters measured by Ocean Optics spectrometer in Figures 9(e) and 9(f), demonstrating that the HSHI system had the capability of capturing hyperspectral images with accurate spectral information. The small discrepancy between the two spectral profiles was mainly due to the light leakages between the circular lightguides in the input end, as discussed in the above section.

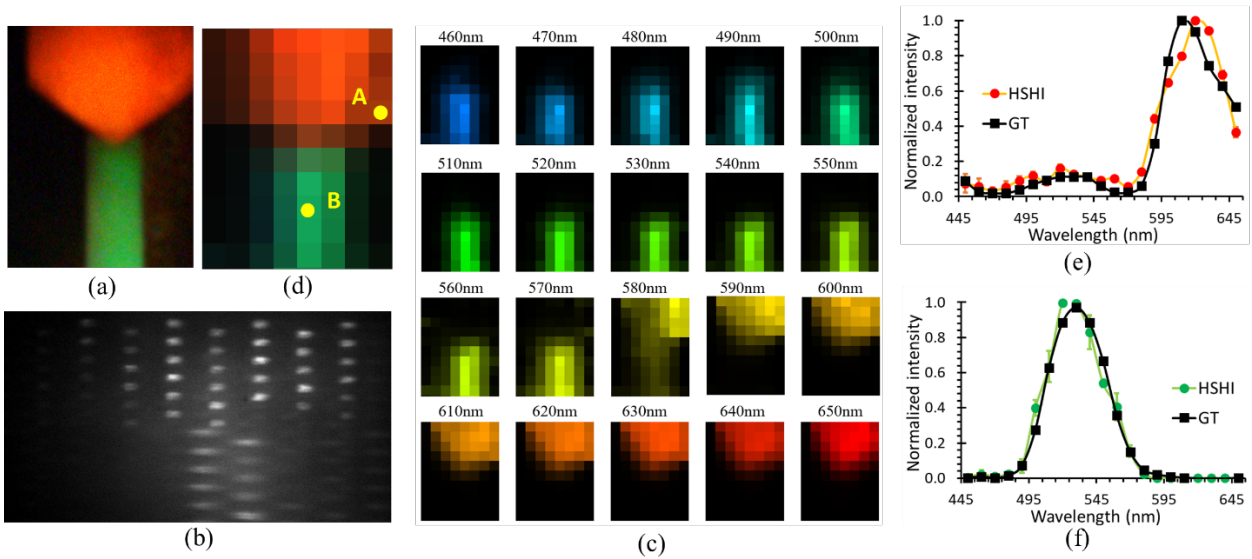

**Figure 9**. Demonstration of the system concept and performance with a manmade flower. (a) Color image of the manmade flower, (b) dispersed image obtained from HSHI system, (b) color image of the manmade flower, (c) hyperspectral images, (d) reconstructed color image from hyperspectral images in (c), (e) normalized spectral profile of the lightguide A in (d), and (f) normalized spectral profile of the lightguide B in (d). The black curves in (e) and (f) are the transmitted spectral profiles (GT) of the color film lighting gel filters used to make the manmade flower.

To demonstrate the potential application of the HSHI system in biology, an experiment was implemented with polytrichum leaf WM prepared microscope slide. Figure 10(a) illustrates the color image of the polytrichum sample, and its down-sampled image is plotted in Figure 10(b). The polytrichum sample showed distinct spatial inhomogeneity. Specifically, the left side of the sample exhibited a more greenish color, while the right side was more oriented to the yellow spectrum. Such inhomogeneity was also captured by the HSHI system and reflected upon the extracted hyperspectral images (Figure 10(d)) and the reconstructed color image (Figure 10(e)). In 460nm and 480nm, the left lightguides gave relatively strong response while in longer wavelength the right lightguides possessed higher intensity, which agreed with the spatial variation of the sample color. Because of the limited number of lightguides, to better evaluate the recovery performance, the color image in Figure 10(a) was down-sampled until it shared the same pixel number as the HSHI system, as shown in Figure 10(b). Under the same pixel scale, the HSHI output possessed a better similarity with the down-sampled image, demonstrating the potential application of the HSHI system in biology. The spectral profiles from the two selected lightguides are plotted in Figure 10(f) and 10(g). The spectral features of point A tended to be inclined to wavelength longer than 570 nm, while the spectral features of point B had a clear blue shift compared with A. It matched the RGB colors of the two points as A is more yellowish and B looks greener.

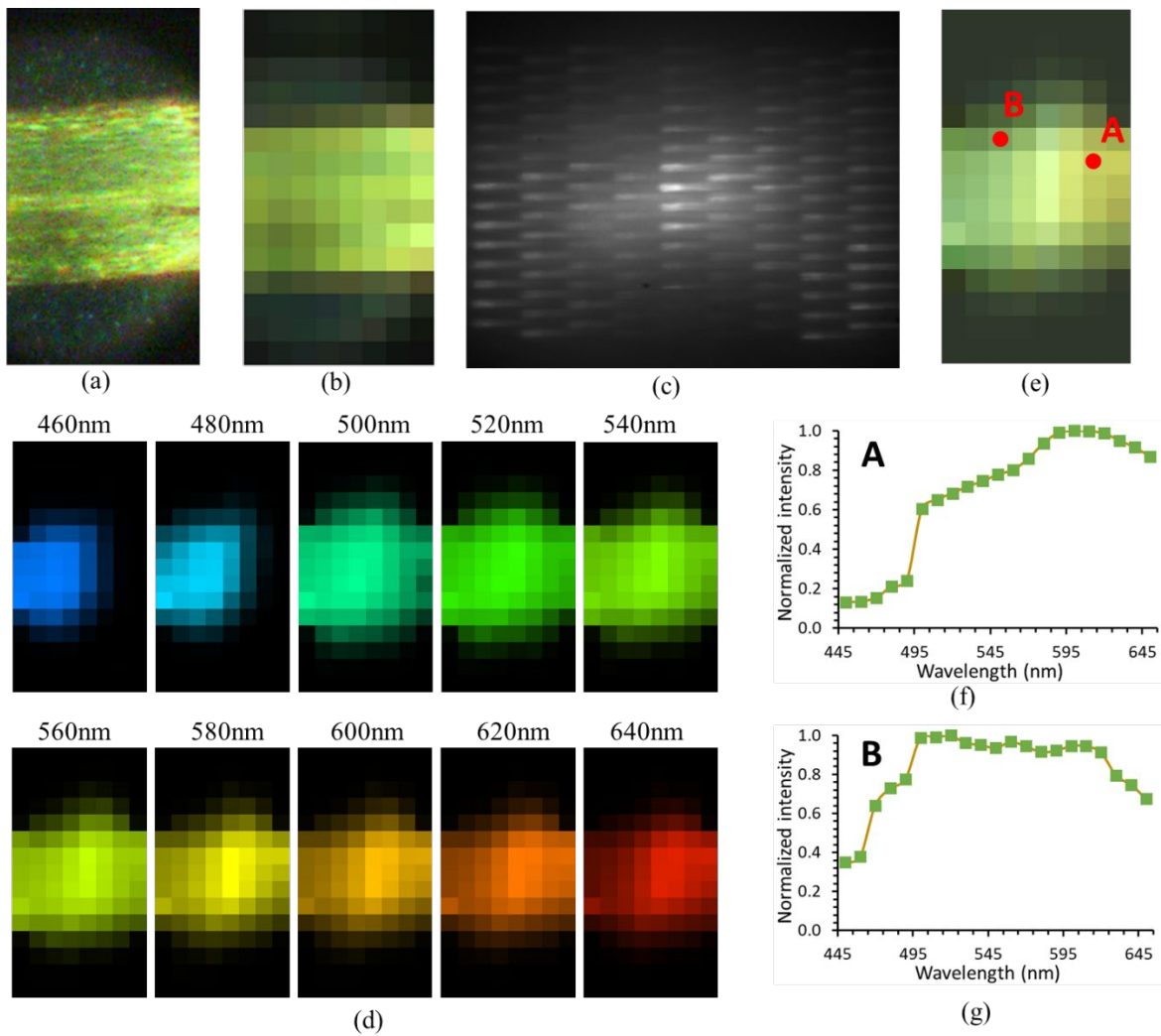

**Figure 10**. Demonstration of the system performance with polytrichum leaf WM prepared microscope slide. (a) Color image of the sample, (b) the down-sampled color image, (c) dispersed image obtained from HSHI system, (d) selected hyperspectral images, (e) reconstructed color image from hyperspectral images, (e) normalized spectral profile of the lightguide A in (e), and (f) normalized spectral profile of the lightguide B in (e).

To demonstrate the snapshot imaging capability, a dandelion fuzz sample in Figure 11(a) was imaged by a dynamically varying illumination. In details, the illumination spectrum switched from 450nm to 650nm sequentially with a bandwidth of 10nm, along with one with 100nm bandwidth centered at 580nm. For each illumination spectrum, a dispersed image was captured from HSHI system, and the same hyperspectral reconstruction operations were applied to extract hypercube and reconstruct color image. To evaluate these outcomes, the ground-truth color images of the dandelion fuzz sample under the same illumination conditions were captured for comparison. The results of total 22 illumination scenarios were recorded to show the dynamic spectral recovery of the proposed HSHI system. Among 22 frames, the results of two snapshots with illumination centered at 510nm and 640nm are displayed in Figures 11(b)-11(e) and Figures 11(f)-11(i), respectively. Figures 11(c) and 11(g) are the unprocessed dispersed image captured from HSHI system. From 510nm to 640nm, the bright spots underwent a leftward shift which agreed with the spectral calibration result. Compared with the ground-truth color images in Figures 11(b) and 11(f), the recovered color image shown in Figures 11(d) and (h) demonstrated accurate spatial distributions and RGB colors, considering the pixelated effect due to the limited number of lightguides. The light spectral profiles for one selected lightguide (the yellow point in Figures 11(d) and 11(h)) under two illumination spectra are plotted in Figures 11(e) and 11(i), showing accurate center wavelength and bandwidth. All the spectral profiles were normalized based on the maximum value within the 22 frames.

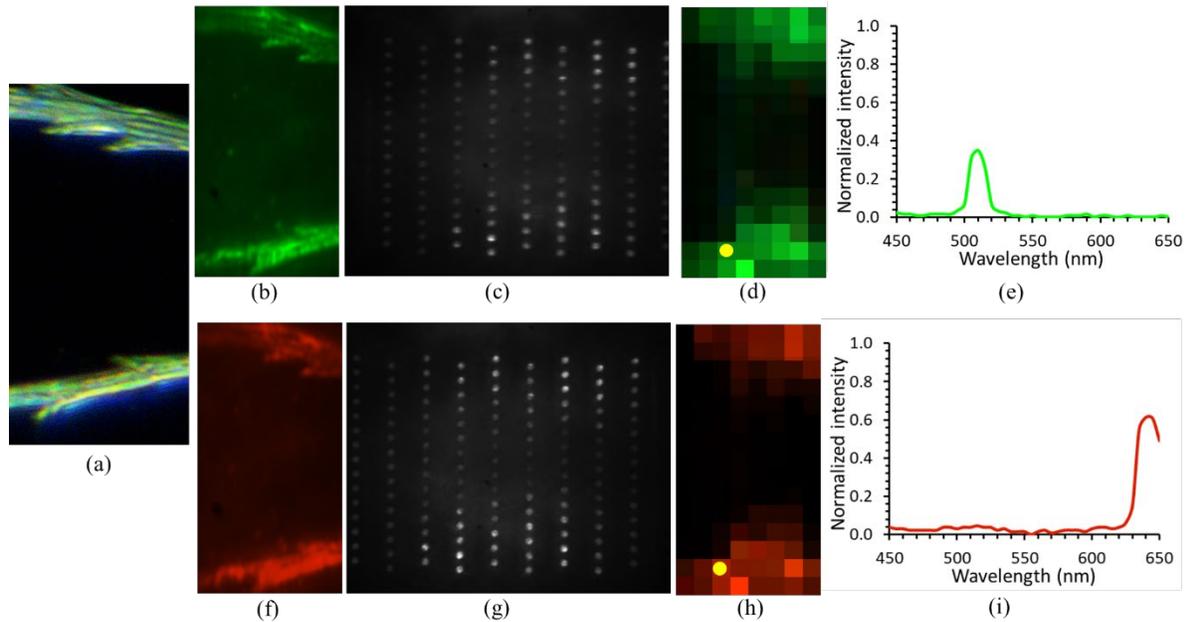

**Figure 11**. Demonstration of the snapshot hyperspectral imaging capability using dandelion fuzz sample. (a) Color image of the sample under visible light illumination, (b) color image of the sample under 510nm spectrum, (c) the dispersed image, (d) the reconstructed color image, (f) color image of the sample under 640nm spectrum, (g) the dispersed image, (h) the reconstructed color image, (e) and (i) are the normalized intensities of the lightguide indicated by the yellow dot in (d) and (h). The intensity was normalized by the highest spectral intensity of the light guide. The video file demonstrates the real time hyperspectral imaging capability. All the spectrum profiles were normalized based on the maximum value within the 22 frames.

**Conclusion and Discussion**

We have developed a compact, high-resolution snapshot hyperspectral imaging technology with 3D printed glass lightguide array. Compared to other methods, the 3D printed lightguide array has several unique advantages as an image sampler: it enables high spatial resolution imaging as the lightguide can be printed as small as the pixel size of the digital sensor; it can dramatically simplify the optical systems before and after the lightguide array as the input and output surfaces of the lightguide array can be curved; and potentially it is low cost as both the new 3D printing technologies and the simplified optical systems can reduce the system cost significantly. In addition, the glass lightguide array will enable the potential applications in UV and NIR, which are not possible with polymer lightguide array.

While we have demonstrated that the 3D TPP printing process is an excellent method for fabricating the lightguide array, some further developments are still needed. The first is to print large format lightguide arrays with much more lightguides, for example more than 640x480 lightguides. Increasing the number of the lightguides will increase the aspect ratio of the lightguide dramatically. While TPP 3D printers with long-range high precision motion stages, such as Nanoscribe's Quantum X Shape and Upnano's NanoOne, can print large structures with high resolution, significant efforts are still needed to optimize the printing process to print the lightguide array as each lightguide has a small dimension to achieve high spatial resolution and it is necessary to maintain TIR condition. Another challenge is the printing material, the cured structure should be strong enough without deformation during the printing process as it floats within the printing material. Therefore, the printing material should be optimized to ensure that the cured structure is strong enough. Since the light propagates through TIR inside the lightguide and there is no cladding layer like optical fiber, any dust on the lightguide surface will break the TIR condition, reducing light efficiency and introducing crosstalk. Therefore, the printed lightguide array should be encapsulated so that no dust will accumulate on the TIR surface of the lightguide.

Our near-term plans are to update the home built TPP printer with a high power 780 nm femtosecond fiber laser and 3-axis high precision stage with 26 mm travel range and 10 nm resolution, optimize pre-condensed liquid silica resin to improve curing speed and improve the strength of the cured structure, and optimize the printing process to print large format lightguide. We will also design custom optical systems to demonstrate the HSHI system with lightguide array which has concave input and output surfaces.

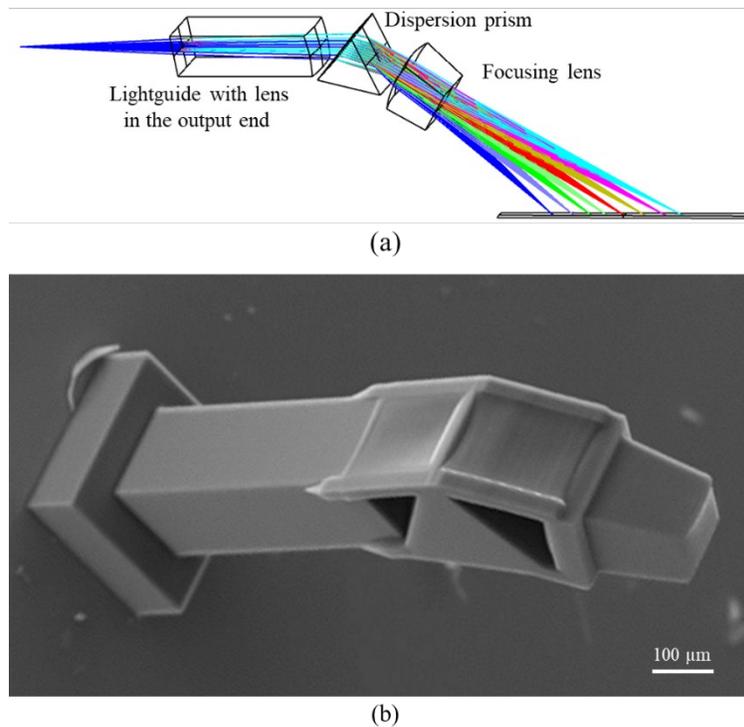

**Figure 12**. Bio-inspired lightguide for hyperspectral imaging. (a) Optical layout of a single lightguide with lens in the output surface, followed by the dispersion prism and focusing lens, (b) SEM image of a 3D printed lightguide with collimation lens, prism and focusing lens.

The proposed concept of lightguide array and the TPP process for printing precision optical elements will address some major limitations in current hyperspectral imaging systems. 3D printing process will not only fabricate individual components, such as lightguide array and lens, it can also fabricate the integrated system without the need of fixtures and active alignment, significantly reducing the cost and system dimension. Figure 12 demonstrates one lightguide with collimation lens, dispersion prism, and focusing lens. It opens new doors for compact HSHI systems, such as the systems in Figures 1(d) and 1(e).

With unique properties in resolution, compactness, and cost, we expect that the HSHI system will have a variety of applications in astronomy, remote sensing, surveillance, biotechnology, medical diagnosis, agriculture, pharmaceuticals, food analysis, oil and gas detection, machine vision, and etc., from UV to IR.

## Funding

## Disclosures

R.L. is the founder of Light Research Inc.

Z. H, Y. S, and P. Y contributed equally to this work. Z. H, P. Y, Y. S, D. L, and R. L conceived the idea and designed the study. P. Y prepared materials and measured the printed samples, Z. H performed printing experiments, and Y. S processed image. Z. H, Y. S, P. Y, D. L, and R. L analyzed and interpreted the results and wrote the manuscript.